\documentstyle[preprint,tighten,12pt,aps]{revtex}

\begin{document}

\input{psfig.sty}

\title{Beyond the Constituent Quark Model
\footnote{Based on a plenary talk presented at Hadrons 2001, Aug 25 - Sept 1, Protvino, Russia.}}

\author{Eric S. Swanson}

\address{
Department of Physics and Astronomy, 
University of Pittsburgh, 
Pittsburgh, PA  15260\\
and Jefferson Lab, 12000 Jefferson Ave,
Newport News, VA 23606}

\maketitle

\begin{abstract}
Modern experiment requires a reliable theoretical framework for low energy
QCD. Some of the requirements for constructing a new model of QCD are presented
here. Progress toward these requirements are highlighted.
\end{abstract}

\section{Introduction}

The constituent quark model has a long and distinguished history of service to 
hadronic physics\cite{cqm}. However, its utility is restricted to quark-number conserving
hadronic processes -- it has nothing to say about channel coupling, or about gluonic
physics in general. And its connection to QCD is tenuous at best. 
Indeed, it is clear that 
present day experiment is outstripping theory and that new reliable and tractable continuum
models of QCD are required to interpret and guide the new generation of hadronic experiments.

For example, it is very likely that resonant structure is seen in the exotic $J^{PC} = 1^{-+}$
channel at 1600 MeV; and something is seen at 1400 MeV at BNL, CERN, and VES. Proving that these
states are mesonic hybrids will require a substantial improvement in our understanding of
the dynamics of soft glue, both in terms of the structure of the putative resonance and in terms
of its coupling to `canonical' mesons. Similarly, it is tempting to interpret the extensive 
Crystal Barrel data on the $f_0(1500)$ as evidence for a scalar glueball. However, state mixing in
the scalar sector is notorious for its strength and its obscurity. This mixing must be 
thoroughly mastered before we can claim the discovery of a glueball and this task will require
a reliable model of soft glue. As a final example, consider the extraction of baryonic resonance
parameters from Jefferson Lab, BNL, GRAAL, and Bonn. At modern energies, one must analyse data
using coupled channel methods; thus $\pi N$, $\pi\pi N$, $\pi\pi\pi N$, $\eta N$, $\rho N$, etc
channels become important and one must have a trustworthy method to parameterise the couplings between
the different channels.
It will also become increasingly important to have reliable estimates of background amplitudes.
A moments' reflection reveals that this is a difficult problem in strong QCD: quark exchange diagrams
contribute, but may also be present at the effective meson-exchange level. More perplexing is the
possibility of quark-antiquark annihilation to intermediate states with excited gluonic content\footnote{There is compelling evidence for this in meson-meson and meson-baryon scattering data\cite{hir}.}.

Although these examples were drawn from the mainstream of hadron spectroscopy, it should be 
stressed that these issues are rather far reaching. For example, extracting electroweak
phases will require reliable knowledge of strong phases which are generated in the necessarily
present hadronic final states.  Analyzing forthcoming RHIC data for putative signals of 
the quark-gluon plasma will require a careful subtraction of hadronic scattering background
which may mask the signal. Again, a thorough knowledge of hadronic dynamics is needed.

Of course, more fundamental reasons exist for building a new model of strong QCD. QCD is a
remarkably rich theory, displaying such diverse phenomena as spontaneous chiral symmetry
breaking, asymptotic freedom, colour confinement, possible new high temperature phases
of matter, and important topological features. It would be remiss to abrogate the
responsibility of learning as much as possible about these phenomena.

\section{Issues Facing a Modern Constituent Quark Model}

A short and incomplete list of the issues facing the construction of a `new' quark model
follow.

\vskip .3 true cm
\noindent
$\bullet\ ${\sl\bf the nature of confinement}

For several decades, the quark model has employed a linear (or similar) static long range 
interquark potential.
While this is in accord with lattice data, many open issues remain. For example, what is the
colour space structure of the long range force? This is required even for heavy quarks. The
standard choice is $\lambda\cdot \lambda$ and this has received strong support from the lattice\cite{casimir}. It is also possible to prove that this is the correct colour structure in the heavy
quark limit\cite{ss7}. The extrapolation to light quark masses remains an open issue. Note that
the colour structure is important when one considers hadronic interactions (ie., processes with
more than three valence quarks).

The Lorentz structure of confinement is not determined by Wilson loop calculations and is 
important for those wishing to `relativise' quark models. Indications from heavy quarkonium
spin splittings and from direct lattice computations are that the structure is {\tt scalar}$\otimes${\tt scalar}\cite{scalar}. However, it should be remembered that this is the form of the {\em effective} interquark interaction once gluons are integrated out of the theory/model. Indeed, 
comparison with QCD in Coulomb gauge shows that the Lorentz structure of confinement is
{\tt vector}$\otimes${\tt vector} in the heavy quark limit\cite{ss5}, and that the effective scalar
interaction arises due to nonperturbative mixing with intermediate hybrids.

\begin{figure}[h]
\hbox to \hsize{\hss\psfig{figure=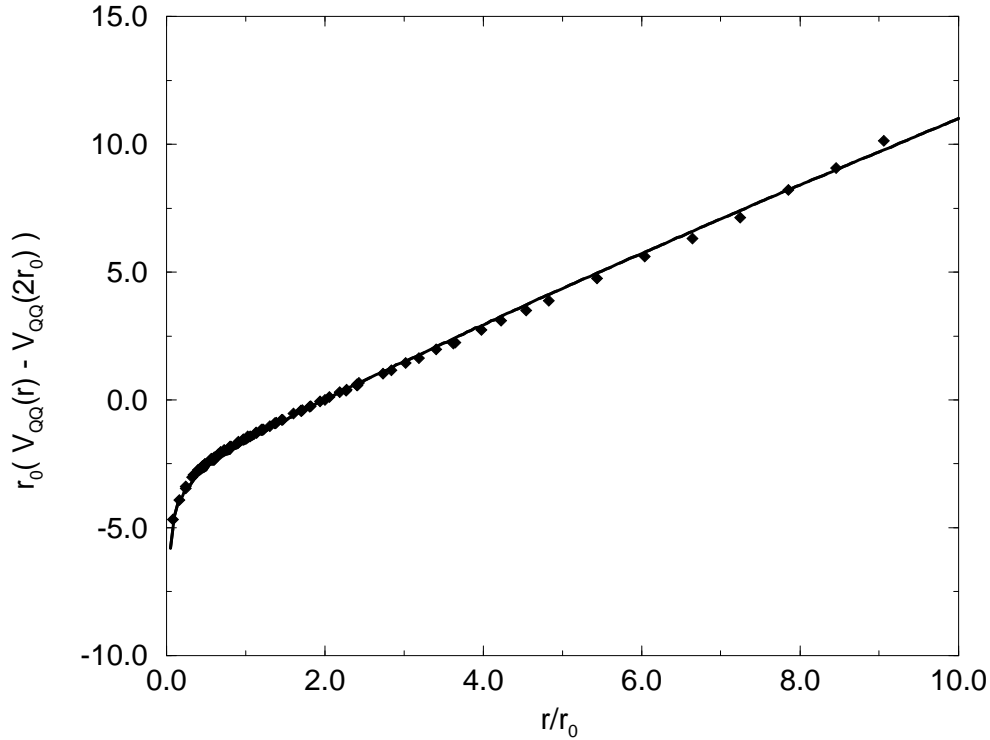,width=5.5in}\hss}
\caption{The leading Tamm-Dancoff Static Potential (line) compared to the
Wilson loop confinement potential (points).}
\end{figure}

An important, and almost uniformly ignored, aspect of model building is deriving the long range
confinement potential. While it is tempting to merely assume that confinement exists and to 
take its form from the lattice, this risks missing vital aspects of strong QCD because
confinement is strongly tied to the vacuum 
structure of QCD, and therefore is related to the appearance of chiral symmetry breaking and
constituent quarks (see `chiral pions' below). All three of these phenomena are central features
of low energy hadronic physics and must be modelled reasonably well before we can have full
confidence in the new quark model.  That this ambitious goal is possible has recently been
demonstrated twice: with the Schwinger Dyson formalism in Landau gauge\cite{axial}, and with Hamiltonian methods in Coulomb gauge\cite{ss7}. The result of the latter calculation is shown in Fig. 1.

\vskip .3 true cm
\noindent
$\bullet\ ${\sl\bf gluodynamics}

Glue, and especially the dynamics of glue, is conspicuously absent from most present
day models of strong QCD. It is clear that glue plays a vital role in many aspects
of low energy hadronic physics. Indeed, about the only place where it is relatively safe 
to neglect QCD gluodynamics is when discussing the static properties of mesons and baryons. 
Thus, for example, a reliable model of gluodynamics is required to address simple questions
such as the masses and static properties of glueballs and hybrids. Just as important is the
way in which these states couple to `canonical' matter. This must be known if we are to 
disentangle exotics from the canonical spectrum. Again the lattice will be of great assistance.
Lattice computations of glueball masses\cite{Morningstar:1999rf} serve as a litmus test for 
any putative models of gluodynamics. Furthermore, high precision computations of the
adiabatic excited gluon energies provide our first glimpse into the dynamics of strongly
interacting glue\cite{JKM}.

\vskip .3 true cm
\noindent
$\bullet\ ${\sl\bf unquenching the quark model}

A closely related issue is going beyond the valence approximation in hadronic phenomenology.
While it is clear that this is a pressing issue for the investigation of nonvalence physics,
such as the strangeness content of the proton\cite{gi}, or the perplexing robustness of the
OZI rule in the face of hadronic loops\cite{gi2}, it is also relevant to spectroscopy. 
Typical hadronic widths of 150 MeV point to typical hadronic mass shifts of a similar scale.
Furthermore, meson loops cause spin splittings which can confound simple attempts at deriving
these. It is clear that a quark model which unifies quark-antiquark pair creation with 
valence physics is required\cite{pz}.

Although there are a number of technical issues which need to be overcome to achieve this
unification (such as efficiently solving coupled channel problems, determining the optimal
number of channels to include in a given problem, and accounting for excluded channels), the
main issue is the form of the quark creation operator. Certainly this operator is dominated 
by nonperturbative glue, but a detailed microscopic description is lacking. The most popular
model to date is the $^3P_0$ model\cite{3p0} (first diagram in Fig. 2) which assumes an 
effective vertex which produces
quark pairs with vacuum quantum numbers. This model produces a reasonably reliable 
phenomenology\cite{3p02}.

Other possible decay mechanisms exist and need to be explored. For example, the 
second diagram in Fig. 2 is the leading diagram in naive perturbation
theory. However predictions of the D/S amplitude ratios (which are sensitive to the 
assumed Lorentz structure of the decay  vertex) in $b_1 \to \omega \pi$ and $a_1 \to 
\rho\pi$ strongly prefer a $^3P_0$ pair creation over 
$^3S_1$\cite{3s1}. 

\begin{figure}[h]
\hbox to \hsize{\hss\psfig{figure=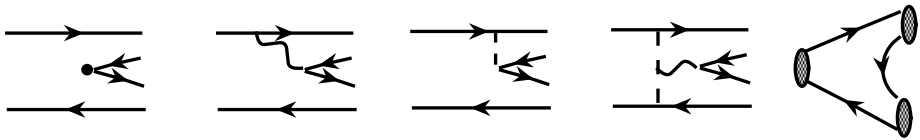,width=5in}\hss}
\caption{Possible decay mechanisms.}
\end{figure}

Another possible decay mechanism is obtained by isolating the instantaneous portion of 
diagram 2 (diagram 3 of Fig. 2) -- say by working in radiation gauge. However, a  model examination
of this process reveals that it is strongly suppressed with respect to the $^3P_0$ 
vertex\cite{Ackleh:1996yt}.

A promising approach to hadronic properties is provided by the Schwinger-Dyson formalism.
The leading decay mechanism  in this approach is the triangle diagram\cite{SDdecay} (last diagram of
Fig. 2). This method has the benefit of employing the same kernel to describe quark-quark
interactions and quark-antiquark pair production. However, this may be too restrictive since
vector pair creation appears to be disfavoured by the $b_1$ and  $a_1$ amplitude ratios. Nevertheless
it is possible that the situation may be saved by the relativistic character of the Schwinger-Dyson
approach (certainly, this physics is not explored in the quark model calculations which
favour the $^3P_0$ mechanism).

The final possibility considered here is the production of a $q\bar q$ pair directly from
the confinement potential/flux tube (fourth diagram, Fig. 2). This diagram is a leading term
in Coulomb gauge QCD\cite{ss7}. Although the phenomenology of decays in Coulomb gauge QCD
has not been explored, it is promising because this diagram yields a vertex with $^3P_0$
quantum numbers, and would be our first microscopic justification of the $^3P_0$ model.

Given the importance of unquenching the quark model and the paucity of our knowledge of the
nonperturbative nature of quark pair creation, 
{\it a lattice exploration of the form of the decay vertex should be a high priority topic
in the near future.}

\vskip .3 true cm
\noindent
$\bullet\ ${\sl\bf chiral pions}

Pions are an important part of any attempt to understand strong QCD. As the lightest hadrons
they dominate nuclear physics; pion cloud effects are important, and pions are ubiquitous
in final states of many hadronic experiments. Having a firm grip on their qualities is of
central importance to constructing viable models of strong QCD. 

It is often said that the constituent quark model view of pions as quark-antiquark
bound states is in conflict with their quasiGoldstone boson nature. However, these two
world-views need not be at odds. For example, the very existence of the constituent quark
model is due to the existence of light pions: the dynamics which causes dynamical symmetry
breaking (and Goldstone bosons) also creates quark-like quasiparticle excitations -- the
constituent quarks\cite{chiral}. A recent paper\cite{ss6}, show explicitly how the Goldstone,
collective, nature of the pion can coexist with the $q\bar q$ bound state quark model pion;
briefly, both descriptions are correct when the appropriate degrees of freedom are
employed (partonic for the Goldstone modes; constituent for the quark model states). Thus it
is likely that a good phenomenology may be obtained simply by ignoring the underlying chiral
aspects of the pion. However, incorporating the physics of chiral symmetry breaking is important
if one wishes to deal with aspects of the QCD vacuum or if the model is strongly constrained 
(so that pionic fluctuations into many-quark Fock components may not be absorbed into 
model parameters).

\vskip .3 true cm
\noindent
$\bullet\ ${\sl\bf relativity}

This is a longstanding and well known problem with the constituent quark model which is
a left-over from the early days of hadronic physics. There is really no reason to continue
with nonrelativistic approaches (except that they are computationally simple and they work
reasonably well!) -- and several groups have mounted efforts to construct `relativised' quark
models. These typically fall into two categories, light cone/Bakamjian Thomas models\cite{lc}
or Schwinger Dyson/Bethe Salpeter models\cite{SD}. The latter are closer to field theory (or 
{\it are} truncated field theory) and offer great hope.  

It is possible to overstate the case for covariance. Any nonperturbative approach must 
break covariance
at some level, for example, the lattice breaks Lorentz invariance by working on a grid and 
models typically must truncate at some level in Fock space. Both of these problems may be
removed in principle -- in practice they are {\it not} removed, but the effects may be checked and
are seen to be small (at least in the case of lattice gauge theory). It is perhaps more useful to 
adopt a practical attitude, for example, it would be useful if  the computation of the
pion decay constant via the PCAC relation $\langle 0 | A_\mu^a(0)|\pi^b(p)\rangle = i f^{ab} p_\mu$
did {\it not} depend on the spacetime index.

\vskip .3 true cm
\noindent
$\bullet\ ${\sl\bf short distance dynamics}

It is easy to believe that the form of the short distance quark interaction is resolved
by QCD; short distance means high $Q^2$, that means small $\alpha_s$, and that means
perturbative one gluon exchange. The phenomenology of one gluon exchange works extremely
well\cite{oge1} and has the virtue of providing a good phenomenology for both mesons and
baryons\cite{oge2}. However the {\it transition}
of one gluon exchange to intermediate or large distance is typically ignored. Indeed, in 
bound state perturbation theory (which is the way all perturbation theory for hadronic 
physics should be performed), the diagram corresponding to one gluon exchange (first diagram
of Fig. 3) corresponds to mixing with intermediate hybrids (second diagram, Fig. 3), and the first diagram does not occur.  However, if one is dealing with a 
field theory, the first diagram reappears as a counterterm which is active
at momentum transfer above the renormalization scale.
How these
two evolve into each other is therefore an issue dealt with by the renormalization group flow of the
underlying field theory and should be properly addressed in a new quark model.  

This issue
is related to a subtlety in most quark models: how are short range and
long range dynamics to be merged? If confinement arises from multiple gluon
exchange, surely it is not correct to simply add one gluon exchange to 
an assumed linear potential.
Addressing this issue is very difficult in covariant gauges, however, in Coulomb
gauge there is a natural separation of instantaneous and transverse potentials
which allows the issue to be resolved simply.

\begin{figure}[h]
\hbox to \hsize{\hss\psfig{figure=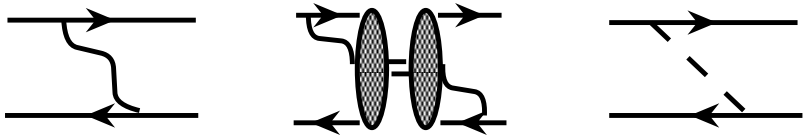}\hss}
\caption{Possible short distance interactions.}
\end{figure}

The last diagram of Fig. 3 represents meson exchange contributions to the quark interaction.
If one admits that pion (and meson) loops can affect hadron properties (as we have argued above)
then one must allow these sort of diagrams. However, it is an open issue as to how important
they are. Robson examined this possibility years ago\cite{DR} in the context of the tensor
splitting of the $S_{11}(1535)$ and $S_{11}(1650)$  and rejected it. It has since been
taken up again\cite{GR}, although not without criticism\cite{Isgur:2000jv}.

\vskip .3 true cm
\noindent
$\bullet\ ${\sl\bf topological aspects}

Shortly after the notion of topology (here we focus on instantons) was introduced to QCD,
't Hooft used instantons to resolve the $U_A(1)$ problem\cite{inst} -- namely that
the axial symmetry of (massless) QCD is not realized in the Wigner-Weyl or Nambu-Goldstone
modes. It has also been argued that collective effects involving infinitely many instantons
may generate the quark condensate, and hence, chiral symmetry breaking. Finally, we mention
that instantons induce an effective quark interaction, however it appears that this force
does not confine\cite{SS}. 

If we are to accept the instanton resolution to the $U_A(1)$ problem, then instanton field
configurations must be accepted as an important subset of the vacuum field configurations,
and their effects should be included in a new quark model. Indeed, computations  with 
instanton models indicate that they may successfully describe many properties of light 
hadrons\cite{SS}.  There is also lattice evidence that instantons dominate the vacuum.
We note, however, that old arguments of Witten against instantons\cite{W} have been
resurrected\cite{Horvath:2001ir}. This paper, in turn, has been criticised\cite{cc}.

There appears to be little room for instantons in the nonrelativistic constituent quark model,
they simply aren't needed to explain the spectrum. However, if instantons do dominate low
energy QCD, they must be incorporated into models. The Bonn group (see Refs. \cite{SD,SDdecay})
has been developing a model which includes instanton-induced quark interactions in a 
relativistic Bethe-Salpeter approach, and the resulting phenomenology appears quite successful. 

Moving beyond the phenomenological stage requires incorporating the effects
of instantons in a way which is consistent with the new quark model's treatment
of the vacuum. And this means that a consistent treatment of confinement,
chiral symmetry breaking, and instanton effects must be found. I know of no
attempts in this direction, and it forms a major challenge for future efforts.

\vskip .3 true cm
\noindent
$\bullet\ ${\sl\bf hadronic interactions}

Hadronic interactions form an important, if under-appreciated, portion of
hadronic physics. They are central to developing a microscopic theory
of nuclear physics, to nuclear astrophysics, to the analysis  of `background'
in $N^*$ and other resonance (hybrids, glueballs) experiments, and to 
electroweak experiments (where hadronic
final state interactions must be properly accounted for). As such, any new quark model
should carry with it a well-defined, tractable, methodology for computing
hadronic interactions. 

The present state of affairs is less than  ideal. Constituent quark model
calculations date  from the '70s\cite{lib}, and continue today with 
resonating group\cite{rg} and perturbative\cite{BS} methods (see Fig. 4). 
While it is
likely that these quark model calculations provide reasonable guidance at
low energies (except for pion-dominated physics where one must hope that
the necessary chiral properties are captured in the quark model -- see the
discussion above), it is less clear how applicable they are at high 
momentum transfer. It is here that light cone approaches\cite{LC} are
expected to be applicable (certainly to inclusive reactions, less certainly
to exclusive). Of course what is needed is a consistent formalism which
allows the computation of hadronic wavefunctions and hadronic scattering
in all energy regimes simultaneously. It is evident that close contact with 
QCD needs to be maintained if these goals are to be achieved.

\begin{figure}[h]
\hbox{
\psfig{figure=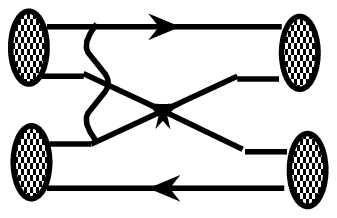}
\qquad\qquad\qquad
\psfig{figure=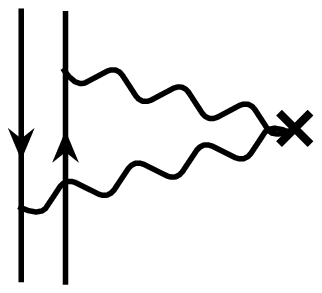}  }
\caption{Hadronic Interactions. The left figure is a basic diagram in
light cone and perturbative quark model computations of hadronic interactions.
It also provides the kernel in resonating group methods. The right figure
is assumed to dominate the interaction of a small meson with an external
colour field.}
\end{figure}

One attempt in this regard was made many years ago by 
Peskin\cite{Bhanot:1979vb}. This approach is essentially a multipole
expansion of the interaction of a small colour singlet state with an external
colour field (see Fig. 4). The resulting dipole interaction is assumed to be
applicable to hadron-hadron interactions as well. However, we note
that one prediction of this model is that the cross section for $\psi'$ with 
hadronic matter is 5000 times larger than that of $\psi$s. Indeed, Peskin 
fears that even the $\Upsilon$ system may be too light for the method to work\cite{p2}.

The historical litmus test for hadronic models has been a computation
of the hadron spectrum.  It is becoming increasingly clear that this is
inadequate because the extraction of resonance parameters is fraught 
with ambiguity. Computations which are closer to the data are
required -- in particular reaction dynamics need to be incorporated
into new quark model predictions. This field is in its infancy; however,
it has started\cite{SL}.

\section{Conclusions}

A crucial aspect of the new quark model is a thorough understanding of the
QCD vacuum. This is required to meet many of the issues raised above:
chiral symmetry breaking, confinement, topology, and gluodynamics. These
issues, in turn, are central to developing a viable model of low energy
QCD.
It will clearly be a stiff challenge to develop a model which adequately
addresses all of these issues; however, hadronic physics provides our
only window into strongly interacting field theory and is a vital
component of nuclear physics, astrophysics, cosmology, and 
physics beyond the standard model\cite{Capstick}. It will therefore
be worth the effort to develop such a model!

An efficient description of hadronic physics will require the identification
of appropriate degrees of freedom -- constituent quarks, massive gluons,
flux tubes, instantons, vortices,  or something new. However, if the ambitious goals
laid out here are to be achieved, a direct connection of these degrees
of freedom to QCD must be maintained.
We can take heart that progress is being made. Of particular note is the
assistance of lattice gauge theory, which promises to be a useful
shortcut to the development of new ideas and to testing these ideas.

\end{document}